\documentclass[conference]{IEEEtran}
\usepackage{multicol}
\usepackage{cite}
\usepackage[dvips]{graphicx}
\usepackage{epsfig}
\graphicspath{{Images/}}
\usepackage{subfigure}
\usepackage{float} 
\usepackage{gensymb}
\usepackage{array}
\usepackage{amsmath}
\usepackage{amsthm}
\usepackage{algorithm}
\usepackage{algorithmic}
\usepackage{mdwmath}
\usepackage{mdwtab}
\usepackage{amsfonts}
\usepackage{psfrag}
\usepackage{textcomp}
\usepackage{psfrag}
\usepackage{color}

\title{Reinforcing Edge Computing with Multipath TCP Enabled Mobile Device Clouds}
 
\author{\IEEEauthorblockN{Venkatraman Balasubramanian, Kees Kroep,  Kishor Chandra Joshi, R. Venkatesha Prasad} 
\IEEEauthorblockA{Delft University of Technology\\
Delft, Netherlands\\
\{v.balasubramanian, H.J.C.Kroep, k.chandra, r.r.venkateshaprasad\}@tudelft.nl}}

\begin{document}
\maketitle
 \begin{abstract}
In  recent years, enormous growth has been witnessed in the computational and storage capabilities of mobile devices. However, much of this computational and storage capabilities are not always fully used. On the other hand, popularity of mobile edge computing which aims to replace the traditional centralized powerful cloud with multiple edge servers is rapidly growing.  In particular, applications having strict latency  requirements can be best served by the mobile edge clouds due to a reduced round-trip delay. In this paper we propose a Multi-Path TCP (MPTCP) enabled mobile device cloud (MDC) as a replacement to the existing TCP based or D2D device cloud techniques, as it effectively makes use of the available bandwidth by providing much higher throughput as well as ensures robust wireless connectivity. We investigate the congestion in mobile-device cloud formation  resulting mainly due to the message passing for service providing nodes at the time of discovery, service continuity and formation of cloud composition. We propose a user space agent called congestion handler that  enable offloading of packets from one sub-flow to the other under link quality constraints. Further, we discuss the benefits of this design and perform preliminary analysis of the system.
 \end{abstract}
 \section{{Introduction}\label{sec:intro}}
 The unprecedented growth in processing and storage capabilities of  mobile devices, warrant new architectural  frameworks for mobile edge computing to utilize the unused resources of mobile devices to support the complex applications~ \cite{mobiforge}.  With an overall growth rate of 29.8\% each year as reported by Cisco \cite{cisco}, by the end of 2020 there would be more than 4.4 billion mobile application users. Out of these, there are around one in four mobile applications that are downloaded once and never used again  primarily  due to the growing application needs that mandate reliance on cloud resources.
 \begin{figure}[ht]
 	\centering
 	\includegraphics[clip,trim=0cm 0cm 0cm 0cm, width=0.8\columnwidth]{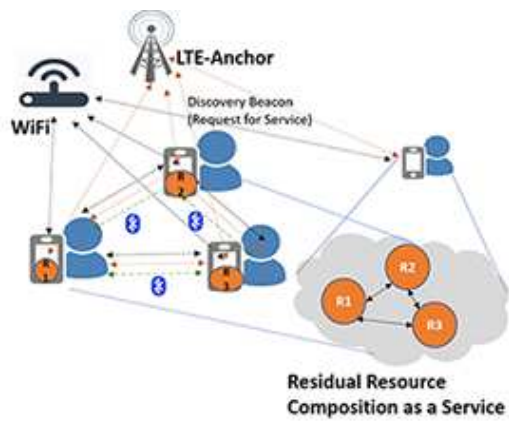}
 	\caption{\textbf{A three unit MPTCP-enabled Mobile Device Cloud -- A service discovery beacon is sent by the consumer by  the available  APs. The APs locate the resources, say three other users who have a residual resource of $R_1, R_2, R_3$. The composition application in the local devices begin resource composition  that together enable an $R2CaaS$.}}
 	\label{fig:main}
 \end{figure}
 If  a  device is able to process part of an application, it is difficult to find the remaining resources to support these  intensive applications vis-a-vis computation  or storage. Thus, many of these applications resort to costly remote cloud services~\cite{Shi:2012:CCC:2342509.2342515, Mtibaa6753815, bala8355198}. Remote cloud services rely on large consolidated data-centers that provide computation capabilities and storage. However, these data centers represent a point of centralization that has serious shortcomings. It can end-up as a single point of failure when applications need real-time services or data center's geographical location is, often times, out of limits for the customers using it.
 Moreover, public clouds have frequent issues such as infrastructure cost and high round trip times (RTT) while considering time sensitive classes of applications. In~\cite{Balasubramanian8072033, McGilvary7214163} authors discuss the services and applications that are infeasible while using remote clouds. As mentioned in \cite{bala8549296}, not all the resources of mobile devices are fully utilized at all times. Therefore, using mobile devices nearby (in place and time) is an attractive alternative to fulfill these requirements. This has led to the concept of using a pool of resources from end-user mobile devices to provide cloud-like services.
 
 This class of cloud computing that deals with the formation and deployment at the user’s level is known as Mobile Device Cloud (MDC). An MDC is a pool of individual devices with limited computational capabilities but which can together form a computing entity closest to the user. This low-cost computational environment is deployed over a network where all nodes cooperatively maintain the network~\cite{Drolia:2015:KKS:2795381.2795388}. Hence, wireless local area networks (WLANs) and Mobile Ad-hoc networks (MANETs) are predominantly considered \cite{Abid:2014:SDR:2658850.2632296}~where users can form a wireless network at any place. A P2P network enabling a computational environment for mobile nodes is another example of an MDC. The primary characteristics of such cloud environments are: (i) both consumer and provider nodes are mobile, and (ii) service composition would change dynamically depending on the available resources (nodes). In this article, we consider these offloading environments where a set of mobile devices collaborate to complete a computational tasks. In the next subsection, we illustrate one such scenario.
 
 \subsection{Scenario}
 With the advent of multiple radio interfaces, a new paradigm to increase the link utilization has been developed which is known as Multipath TCP (MPTCP)~\cite{Paasch:2012:EMH:2342468.2342476}. It can greatly increase application performance using multiple available  paths. As seen in Figure \ref{fig:main}, an added complexity is observed when devices with multiple interfaces collaborate to increase the QoS offered by an MDC (like a crowded stadium  where multiple devices can collaborate to form an MDC). We propose such a scenario where MPTCP enabled devices together form an MDC. Here, a request beacon is sent to an access point for discovering nearby devices who offer their resources for MDC formation. Once the request is received, the access point (AP) will assist in finding the resources.  Once received, the diverse collection of resources (natively stored, and are called  virtual resources) are composed into a usable device cloud infrastructure via service APIs as seen in \cite{Balasubramanian7868393}. The composition algorithm performs the key functions pertaining to the services offered and provided to the consumer. However, due to dynamic link conditions (wireless bottlenecks), frequent link breaks can occur. Further, due to the unstable links,   heavy message passing is invoked,  and link congestion becomes inevitable. We elaborate this problem as follows. 
 \subsection{Problem}
 Devices that form the MDC are far more susceptible to breaking links or compositions due to movements from region to region or excessive load as described in \cite{Mtibaa6753815}\cite{Balasubramanian7868393}. Further, the number of control messages that are being sent between the devices that make use of the same link can overload the network so much so that the multiple re-transmissions  affect network performance adversely. As shown in \cite{Nikravesh:2016:IUM:2973750.2973769} the current development of MPTCP and the reasons for its sub-optimal performance in mobile environments needs further attention. We identify one key issue \textit{i.e.,} delay between multiple flows that causes congestion due to increased buffering in the local device. Hence, our focus is on congestion control.
 \begin{enumerate}
 	\item Congestion Control at individual device: At each individual device there should be necessary packet reordering mechanisms that would allow each device buffer to recognize the sender for correct packet reassembly. If one device has not been upgraded or has not switched to the $MPTCP\_CAPABLE()$ mode then there would be a detrimental impact on the service provided by such a cloud. In summary, excessive on-device buffering (on-device buffer bloat) reduces the collective quality of the Mobile Device Cloud. Hence, our goal is to minimize congestion by facilitating the on-device queues via a user space agent. That in turn reduces the end-end delay.
 \end{enumerate}
 
 \subsection{Contributions}
 \begin{enumerate}
 	\item We propose a novel Multipath TCP enable MDC framework accompanied with  a congestion handler to facilitate an efficient offloading of packets among multiple radio interfaces.
 	\item We define and propose a mathematical modelling of Potential $P_{i,k}$, that measures the ability of the device to serve requests successfully in an MDC. Further, we investigate the effects of small sized queuing to facilitate traffic between the individual resource entities in a mobile device pool. Essentially, we minimize the on-device buffering that affects overall delay in servicing the request.
 	\item A proof of concept preliminary experiment is performed that demonstrates the novelty of the design, by using a synthetically generated \textit{Nakagami-m} distribution model to generate the random realizations of wireless channel. A preliminary analysis shows that the proposed congestion handler can assist in mitigating congestion delay under the channel fluctuation conditions.
 \end{enumerate}
 The remainder of this paper is arranged to delineate the state-of-the-art approaches in Section \ref{sec:relate} followed by the concepts related to our design in Section \ref{sec:concepts}. In Section \ref{sec:system} a mathematical formulation to manage congestion and improve quality of service is proposed. Finally, in Section \ref{sec:eval} we take some measurements and discuss the proof of concept experiments, followed by conclusion remarks in Section \ref{sec:conclusion}.

\section{{Related Works}\label{sec:relate}}
In \cite{Balasubramanian8072033}, the importance and need of a Mobile Ad-hoc Network (MANET) cloud formation in MANETs is highlighted. It shows how the mobile ad-hoc composition is formed, managed and utilized. However, the impact of  message passing and congestion probabilities among the devices in the composed environment is not considered. 
In \cite{Paasch:2012:EMH:2342468.2342476} authors show how the MPTCP coupled congestion algorithms are designed with an implicit assumption of users experiencing the same amount of losses at shared bottlenecks.
In \cite{Shi:2012:CCC:2342509.2342515},   a clinical approach towards design of a device cloud is proposed.  In \cite{Mtibaa6753815} an addition to this approach is proposed that support this design with an approach where the computational loads and its effect on  individual devices is measured. Likewise, \cite{Ho8254706} presents a mobile inter-cloud architecture that takes assistance from virtualized mobile terminals in a user specified application environment. 
In \cite{Baskett6488568}  a software defined network (SDN) enabled Ad-hoc cloud formation relying on the presence of a global controller is proposed. Authors in \cite{Popovici:2016:EMC:3005745.3005769} realize a congestion controller based on policy drop method that falls under a game theoretic approach. An artificial loss is added to all TCP and MPTCP traffic flows based on the link speed.

The above designs primarily consider a single path TCP approach or a base station controlled D2D strategy for communications among the mobile device cloud members. To a large extent the MPTCP approaches are hugely prevalent in device to server networks than among the D2D networks. Our approach differs from all the above approaches in considering an MPTCP enabled device cloud where each end point performs user space congestion control that performs better than traditional kernel implementation.  

\section{Concepts}  \label{sec:concepts}
\subsection{Device Cloud Composition}
In a mobile device cloud, due to the resource limitations in a device, computation-intensive applications require external assistance for execution. For instance, the tasks which cannot be processed locally and require a resource rich environment would need to be offloaded to the cloud service providers. As the centralized public clouds are either inaccessible or require a higher round trip time, in time-bound task execution scenarios (like application execution in music concert in a crowded stadium) it becomes infeasible. Hence, the consumer makes a request to the closest peers or those who are ready to collaborate. Considering scenarios such as those with a high density of user-device cloud service providers (DCSP) these requests submitted by the consumers are exposed to a larger set of  DCSP as observed in \cite{Balasubramanian7868393,Balasubramanian8072033, Mtibaa6753815}. In doing so, each device in a composition behaves in server-mode while providing a service and in client-mode while utilizing the service. At no given time can a DCSP have both modes on. Likewise, if all the interfaces are busy, that particular DCSP cannot participate in another collaboration.

\subsection{Multipath TCP}
MPTCP \cite{Paasch:2012:EMH:2342468.2342476} is an extension of the TCP that enables multiple interfaces to engage in a TCP connection in a client/ or a server. MPTCP spreads the traffic over multiple interfaces and splits the application data streams into sub-flows that traverse through the multiple links. In doing so MPTCP ensures maximization of bandwidth and all available resources. Each interface corresponds to one sub-flow to increase the throughput. For instance, in a mobile environment each device will have a cellular, WiFi interface along-with a Bluetooth or Zigbee module. Such devices can make use of MPTCP by preserving connections while moving around. For example, as seen in Figure~\ref{fig:main} a set of users connected to a WiFi and LTE access points can simultaneously leverage the network bandwidth (even if they continue to maintain the Bluetooth links for file transfers). Nevertheless, with MPTCP, changing IP address (due to movements between heterogeneous networks) does not require new session establishments as seen in traditional TCP connection disruptions. Each MPTCP connection is associated with a set of TCP sub-flows that are underlying. These sub-flows can be removed without affecting the application. In the next sub-section we see how the three-way token exchanges and the final ACK are maintained by AP at the time of composition.
\begin{figure}[ht]
	\centering
	\includegraphics[clip,trim=0cm 0cm 0cm 0cm, width=0.5\columnwidth]{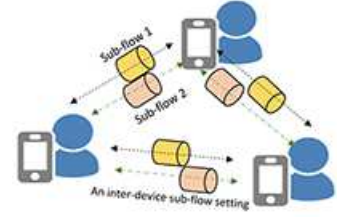}
	\caption{Inter-Mobile Server-Mode Communication}
	\label{fig:server}
\end{figure}

\subsection{Residual Resource Composition as a Service: An MPTCP Device Cloud}
Each MPTCP enabled device that participates in a cloud formation together provides a service called the Residual Resource Composition as a Service (R2CaaS). Consider a crowded stadium where a music concert is being played. The network in the stadium can have WiFi, LTE support other than technologies like Bluetooth. A user who wishes to process an application like music OCR that converts real time voice to lyrics requires computation abilities closest to him as requesting a cloud service would mean a larger round trip time. In such time bound scenarios, a request to the access point is sent to discover the closest devices whose residual resources (intra-device resources that remain idle) are volunteered. As seen in \cite{Balasubramanian7868393}, a cloud composition client runs that homogenizes all the disparately available resources which are made available in usable form to the consumer. Figure \ref{fig:server} explains server mode communication of the system.

\section{System Design}\label{sec:system}
We design our system in order to increase the overall requests accepted by a complete resource composition. Once a request is received by a composition, it means that each individual entity has agreed to participate in providing Residual Resource Composition as a Service (R2CaaS). Our primary aim here is to maximize the requests accepted while maintaining device side congestion control. There is always a constant surge of messages when a device participates in a composition in terms of discovery, session continuity, task execution deadline updates, to name a few. These packets can exponentially grow if the number of devices go on increasing in a composition. Let the $i^{th}$  device has a residual resource of $r_i$.  Therefore, the total combined resource of such a composition, where $n$ is the total number of devices in the device cloud can be expressed as,
\begin{equation}
R= \Sigma_{i=0}^{n} r_i.
\end{equation}
This is the resource granted from a particular MDC.  Let $r_{r}$ be the total requests received and $r_{s}$ units be the total requests serviced since the inception of a particular R2CaaS. We define potential  of the composition $P_{mdc}$ as, the ability of a device to serve requests successfully
\begin{equation}
P_{mdc} = \frac{\Sigma_{i=1}^{n}\Sigma_{k=1}^{m}{r_{s}(i,k)}}{\Sigma_{i=1}^{n}\Sigma_{k=1}^{m}{r_{r}(i,k)}}.    
\end{equation}
The denominator denotes the total requests received since the beginning of the formation of an R2CaaS.  This also includes the failed tasks, i.e., the requests that could not be served within the desired deadlines. The numerator denotes the total number of successfully serviced tasks.
Further, going down to the device level, we decompose the contribution of individual devices to  $P_{mdc}$  in a pool of $n$ devices as given below,
\begin{equation}
P_{i,k}= \frac{{r_{s}(i,k)}}{{r_{r}(i,k)}}.
\end{equation}

A threshold for each link\footnote{In this paper we use the terms "interface" and "link" interchangeably.} $\tau_i$ defines the maximum allowable queuing before the link gets congested. Beyond this value the link would start dropping packets. 
We define $q_i$ as representing all the queued requests as a single row matrix,
\[
q_{i,k}=
\begin{bmatrix}
q_{1,1}, q_{1,2}, q_{1,3}..q_{i,m}
\end{bmatrix}
\]
\noindent $\forall k \in $ \textit{queued  requests} pertaining to an interface. Each interface is defined by link quality indicator $w$. This is introduced based on constant sniffing of the fluctuating link. We assign $w_1,w_2,w_3..w_n$ for different interfaces indicating link quality.
Our aim is to evaluate the number of accumulated requests on a particular interface within $\tau_i$,
\begin{equation} \label{eq:1}
\tau_i \leq q_i w_n 
\end{equation}

\textbf{Why should the queues of interfaces have a threshold?} As observed before our aim essentially is to maintain the Potential ($P_{mdc}$) of an MDC. What comes as a trade-off here is the number of requests that can be handled at one point in time while the local device is continuing to function. The main purpose of using this technique is : \textit{Requests need to be serviced at reduced times}. So, what we are interested is \textit{how long} each request takes in the queue more than \textit{how many} requests can sit inside the queue. It has been proven by literature \cite{Van,Nichols:2012:CQD:2208917.2209336} that continuing to fill the buffer would lead to higher latency, but, discarding packets would lead to reduced QoS. To this end, our endeavour is to offload packets to a corresponding interface that is continuing to function when the MDC is in operation. So, in a device each interface that has a corresponding queue is given a weight such that this threshold is maintained and any other packet entering (based on a Poisson arrival $\lambda$) above this $\tau_i$ is offloaded to another queue. The case of all queues reaching their respective $\tau$ values is beyond the scope of this research. 

\textbf{Modelling $\tau$ by taking assistance from $Nakagami-m$ distribution}:
The Nakagami-m model encompasses a large class of channel variations \cite{liu1427704}. Hence, for our test-case analysis we make use of this model. Based on the fading parameter $m$, of the Nakagami-m model we decide packet movements between queues. We propose a scheme such that the channel fluctuations decreases from 0 to 1. As given by the Nakagami-m model, we assume the fading parameter $m$ ranging from 0 to 1 as high fading regions. That is, m=0.5 is the mid range and m =1 Rayleigh distribution according to \cite{liu1427704}. Our motivation towards applying a one to one mapping to the fading parameter $m$ can be seen from the modelling below.
We know from equation \ref{eq:1} that
\begin{equation}
\tau \leq b_i w_n
\end{equation}
Now Nakagami model provides a link fading parameter as $m$. Intuitively, the $w_n$ parameter measures the link weights so approximately values of $m$ and $w$ have to have a one to one mapping. As shown below we have a proportionality given as,
\begin{equation}
w \alpha m
\end{equation}
\begin{equation} \label{eq:resub}
w = km
\end{equation}
Here, $k$ is the constant of proportionality. Further re-substituting equation \ref{eq:resub} in \ref{eq:1} we have,
\begin{equation}
\tau \leq k b_i m
\end{equation}
We assume fluctuations to be nil when m=1.
Further as seen in Figure \ref{fig:m1}, we set m=0.1-0.5 as the worst case (high fading regions) and progressively as we approach m=1 fading reduces. In section \ref{sec:eval} we make comparisons with non congestion-handler scenario and provide simulation measurements.

\subsection{Case of Forced Drop: Congestion Handler}
The service requests at the AP are purely based on the volunteers who register themselves for R2CaaS services. These volunteers do not guarantee a specific capacity for processing or storage but guarantee participation only if they are willing to complete the requests. That is, all the devices registering to an Access Point would lead to AP breaking down or congestion at the AP. Instead consider a scenario, where only the friends who know each other will allow their fellow friends to be serviced first, the requests that come later need not be served as those requests would be coming from other attendees in the stadium. In such cases, requests are forcefully dropped to maintain MDC operation, reduce congestion and maintain consistent service. 

\begin{figure}[ht]
	\centering
	\includegraphics[clip,trim=0cm 0cm 0cm 0cm, width=\columnwidth]{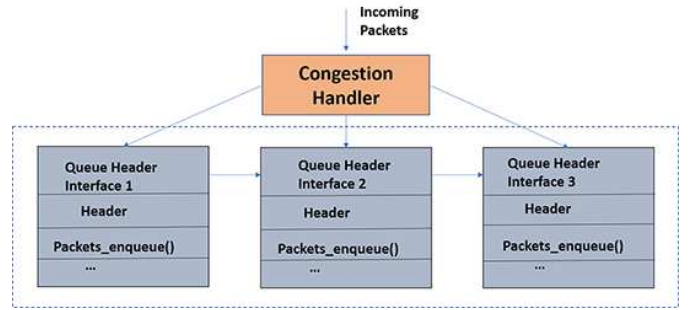}
	\caption{Working of Congestion Handler}
	\label{fig:queue}
\end{figure}

\subsection{Functional Interaction}
\begin{enumerate}
	\item We are interested to de-congest an existing interface queue by a user-space entity which is the congestion handler. When the packets arrive consider the native buffer size to be 20 packets/queue, say the threshold is 15, this is done to keep a low latency while serving the requests at the same time maintain a consistent flow of packets with some requests being offloaded. Assuming there is no forced dropped.
	\item If there is an existing congestion, there are two factors which are checked. Firstly, the number of re-transmission packets which needs to be sent. These packets are pushed into another interface queue. If re-transmission timer times out, then TCP re-intiatilizes window size to 1. Typically, slow-start phase begins again. Secondly, if an interface has failed continuously, then all of the packets queued from the point of failure detection are pushed into another interface queue.
	\item Each of the above operations are based on the congestion handlers calls (updating $\tau$) that are obtained based on sniffing the busy interfaces and the queues.
\end{enumerate}

\begin{figure}[ht]
	\centering
	\includegraphics[clip,trim=0cm 0cm 0cm 0cm, width=\columnwidth]{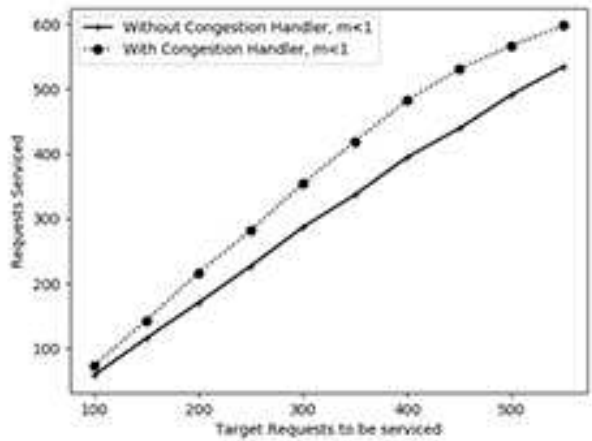}
	\caption{Requests serviced}
	\label{fig:naku}
\end{figure}
\section{Preliminary Evaluation and Experiment}\label{sec:eval}
In order to evaluate traffic characteristics and the network environment we consider a custom topology with $m$ as the Nakagami fading factor. We perform simulation in Python and utilize the probability density function obtained as the basis for channel variation. A uniform distribution of m parameters from 0.1-1 (as values greater than 1 is assumed to be a nil fluctuation) is taken to emulate a randomly changing wireless environment based on which the Congestion Handler functioning is evaluated.  We assume high-fluctuation conditions. In conditions of high-fluctuation, there occurs the main job of the congestion handler to move packets between the queues and not dropping them until congestion is removed (e.g, enough availability of bandwidth). 
We consider aggregation queues to be a size that of 100 packets/aggr.queue and increase till 1000 packets/aggr.queue. We take an average of 10 iterations and plot the graphs. For individual queues, one can divide the number of devices in an MDC and make an approximate calculation. However, we avoid that to cater to the scope of this work. A uniformly distributed $m$ values ranging between 0.1-1 is taken that is set as our region for queue transitions with respect to fluctuations. Through this experiment the goal is to show how channel quality fluctuations can affect queuing behaviour. Our intention is to reduce the packet drops in comparison with the non-congestion handler scenario. Hence, we are concerned only with the $m$ parameter and use the uniform distribution obtained for toggling between queues.
\subsection{Observations}
\begin{figure}[ht]
	\centering
	\includegraphics[clip,trim=0cm 0cm 0cm 0cm, width=\columnwidth]{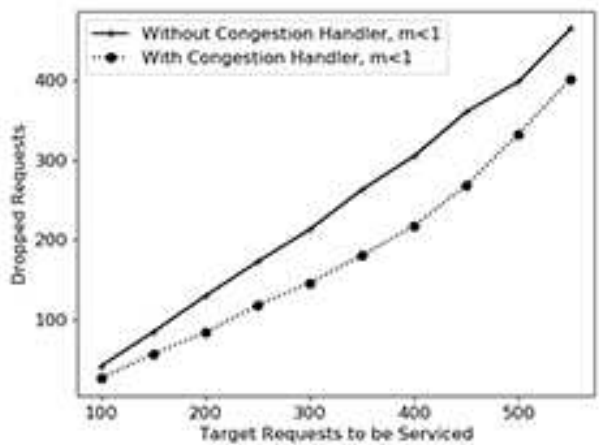}
	\caption{Requests Dropped}
	\label{fig:m1}
\end{figure}

\begin{enumerate}
	\item Percentage improvement- We observe in Fig.\ref{fig:naku} a remarkable percentage improvement using congestion handler. Over 63\% improvement is observed requests serviced with congestion handler minus the request serviced without the congestion handler/100*requests serviced with congestion handler.
	\item Constancy- As seen in the figure \ref{fig:m1}, the dropping of requests reaches a constant low for the case of congestion handler after a certain stage unlike the non-congestion handler case where the drop remains high. Hence, our objective is achieved.
\end{enumerate}
\begin{figure}[ht]
	\centering
	\includegraphics[clip,trim=0cm 0cm 0cm 0cm, width=\columnwidth]{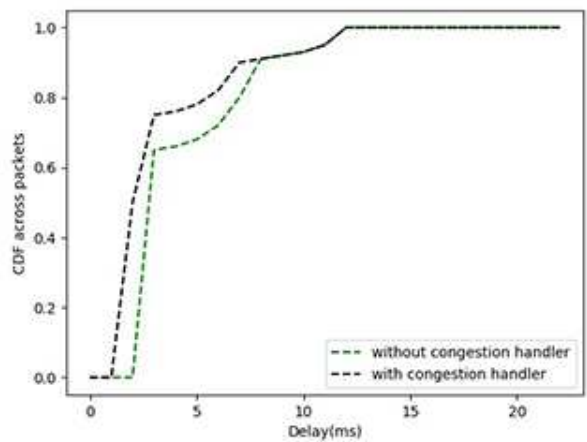}
	\caption{CDF comparison of request completion times}
	\label{cdf}
\end{figure}

\subsection{Discussion and Steps-forward}
Having a user space agent does not always show a positive performance benefit. In Figure \ref{cdf} the request sizes are kept above 1MB in size. While having a requests sizes less than that a negative performance is observed. This can be attributed to the need for further investigation. As future steps the kernel implementation of Multi-path TCP v0.94 stable release needs to be used. Measurement of delay by sending ICMP sequences and the evolution between the time-stamps gives us the variation in delay. Additionally, a thorough network analysis with different interfaces to evaluate the bandwidth and latency issues faced while a device cloud is formed and the corresponding congestion issues faced in due course will be evaluated in future.

\section{Conclusion \& Future Work} \label{sec:conclusion}
In this work, we have proposed an MPTCP enabled Mobile Device Cloud (MDC) and investigated the effects of wireless channel on the scheduling process of the proposed system. We show that how  the congestion handling in MDC environments is integral for maintaining QoS. We demonstrate that the effective use of multiple paths in a dynamically changing environment has its own benefits like fast service times, higher bandwidth to name a few. Due to the inexpensive nature of MDC computation environments, it becomes a plausible choice in time-bound situations where the costly remote cloud services cannot be of much use. The use of MPTCP protocol enables resource rich environment that can further enhance the benefits of an MDC.
We produce preliminary analysis of the proposed Congestion Handler through our modelling of the link quality constraints based queuing in a highly mobile environment with multiple radio interfaces. 
As we are moving closer to realizing an ``Edge Cloud'' for faster computation and low-communication latency, user-defined cloud formation would play a major role in providing on-demand services. In future, we plan to consider scheduling issues in device clouds with MPTCP kernel enabled algorithms, where the control is moved to the MPTCP stack. In doing so, it would be worthwhile to analyze how the network parameters are affected in such environments.
\section*{Acknowledgement}This work is supported by SCOTT http://www.scott-project.eu which has received funding from the Electronic Component Systems for European Leadership Joint Undertaking under grant agreement No 737422. This joint undertaking receives support from the European Unions Horizon 2020 research and innovation program and Austria, Spain, Finland, Ireland, Sweden, Germany, Poland, Portugal, Netherlands, Belgium, Norway.
\bibliographystyle{IEEEtran}
\bibliography{RefMEC}

\end{document}